\documentclass[]{aa}
\usepackage{graphicx}
\usepackage{enumerate}
\usepackage{natbib}
\usepackage{txfonts}
\usepackage{verbatim}
\usepackage{xcolor}
\bibpunct{(}{)}{;}{a}{}{,}    
\usepackage[colorlinks,linkcolor=blue,citecolor=blue,linktocpage=true,breaklinks,plainpages=false,urlcolor=blue]{hyperref}

\begin{document}

\title{Three-dimensional non-LTE radiative transfer effects in \ion{Fe}{i} lines}
\subtitle{IV. Line formation at high spatial resolution}

\author{R.~Holzreuter\inst{1,2}, H. N. Smitha\inst{1}, S.~K.~Solanki\inst{1,3} }

\institute{ Max-Planck-Institut f\"ur  Sonnensystemforschung, 
Justus-von-Liebig-Weg 3, 37077 G\"ottingen, Germany\and 
  Institute of Astronomy, ETH Zentrum, CH-8092 Zurich, Switzerland \and 
  School of Space Research, Kyung Hee University, Yongin, Gyeonggi 446-701, Republic of Korea\\
  \email{holzreuter@mps.mpg.de, narayanamurthy@mps.mpg.de, solanki@mps.mpg.de}}

\offprints{R.~Holzreuter}

\date{Received $<$date$>$; accepted $<$date$>$}

\abstract
{In the first three papers of this series, we investigated the formation of photospheric neutral iron lines in different atmospheres ranging from idealised flux tube models to complex three-dimensional magnetohydrodynamic (3D MHD) simulations. The overarching goal was to understand the role of Non-Local Thermodynamic Equilibrium (NLTE) and horizontal radiative transfer (RT) effects in the formation of these lines.}
{In the present paper, we extend this investigation using a high resolution MHD simulation, with a grid spacing much smaller than the currently resolvable scales by telescopes. We aim to understand whether the horizontal RT effects imposes an intrinsic limit on the small scale structures that can be observed by telescopes, by spatially smearing out these structures in the solar atmosphere.}
{We synthesize the Stokes profiles of two iron line pairs, one at 525 nm and other at 630 nm in 3-D NLTE. We compare our results with those in previous papers and check the impact of horizontal transfer on the quality of the images.}
{Our results with the high resolution simulations align with  those inferred from lower resolution simulations in the previous papers of this series. The spatial smearing due to horizontal RT, although present, is quite small. The degradation caused by the point spread function of a telescope is much stronger.}
{In the photospheric layers, we do not see an image degradation caused by horizontal RT that is large enough to smear out the small scale structures in the simulation box. The current generation telescopes with spatial resolutions smaller than the horizontal photon mean free path should in principle be able to observe the small scale structures, at least in the photosphere.}


\maketitle
\section{Introduction}\label{sec:fluxsheet_intro}
The new generation of solar telescopes, such as the  Daniel K. Inouye Solar Telescope \citep[DKIST,][]{Rimmele2020} in Hawaii or the upcoming  European Solar Telescope \citep[EST,][]{quinteroNoda2022} on the Canary Islands, help us observe the dynamics in the solar atmosphere at unprecedented spatial, spectral and temporal scales. 
In particular, these telescopes can in principle resolve structures well below the horizontal photon mean free path of 50-100 km in the solar photosphere under the assumption that the solar atmosphere is plane-parallel \citep{hinode2019}. However, observations taken at the diffraction limits of these telescopes may capture the imprints of horizontal radiative transfer (RT) or the three-dimensional (3-D) RT effects.  

Atmospheric structures (in density and temperature) at spatial scales well below the typical photon mean free path $\lambda$ may be produced by magneto-hydrodynamic (MHD) processes, in particular by turbulence and magnetoturbulence. However, whether these structures can be observed has been a matter of debate. A row of investigations with two dimensional (\mbox{2-D}) idealized arrangements of the atmospheric structures has been performed \citep[see, e.g.,][ and references therein]{stenholmstenflo1977, stenholmstenflo1978, kneer1981, brulsvdluehe2001, holzreutersolanki2012}. While these papers all considered the \ion{Fe}{I} lines at 525 nm, \cite{kneer1981} looked into the horizontal transfer effects in the solar chromosphere.  These investigations have shown that the horizontal radiative transfer (RT) in combination with strong line scattering may lead to a spatial smearing of the atmospheric structures and, therefore, to a reduced contrast in the observations. They also found that in the photosphere, the spatial smearing due to horizontal RT is insignificant, and they are more dominant at chromospheric heights where lines are strongly scattering \citep{kneer1981}.

More recently, \citet{judgeetal2015} reconciled the different results by finding that the amount of scattering in a specific atomic line is the main responsible ingredient determining the degree of spatial smearing due to NLTE effects, especially in the chromosphere. The authors predicted that this type of smearing results in a solar "fog" whose effects are comparable to that of the point spread function (PSF) of a telescope. The chromospheric spectral lines are well known to be affected by the 3-D non-Local Thermodynamic (NLTE) effects  \citep[for e.g.,][]{leenaartsetal2012, bjorgenetal2018, bjorgenetal2019, judge2020}. However, to our knowledge no study has been done that tests how clearly the new telescopes will be able to see structures at scales down to 20-30 km in the solar photosphere using high resolution state-of-the art 3D MHD simulations. The question is particularly acute for the magnetic field, which is typically measured in lines of \ion{Fe}{I} that are affected by departures from Local Thermodynamic Equilibrium (LTE) \citep[e.g.]{solankisteenbock1988, smithaetal2020a, smithaetal2023}.

\begin{figure}
\centering
    \includegraphics[width=0.45\textwidth]{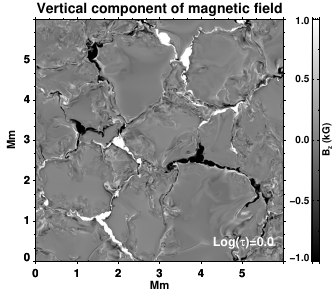}
    \caption{A map of the vertical component of the magnetic field at the surface ($\log (\tau_c^{500}) =0.0$).}
    \label{Figure:fig1}
\end{figure}

In our previous work \citep[][ papers I, II, and, III hereafter]{holzreutersolanki2012, holzreutersolanki2013, holzreutersolanki2015} on photospheric \ion{Fe}{i} lines, we found that many atmospheric structures of size smaller than $\lambda$ remained observable, confirming earlier investigations \citep[e.g.][]{brulsvdluehe2001} on that topic. In papers II and III, where we used radiation-MHD simulations to model the solar photosphere, we also found, that the results of the idealized models of flux tubes (FT) and flux sheets (FS) do not correspond well to those obtained using the  much more realistic MHD simulations. The main shortcoming of these earlier studies was that the MHD simulations employed had a grid spacing of 20 km in the horizontal direction, i.e. too large to address the kind of resolution that is expected to be achieved by the new telescopes. Therefore, in this work, we extend these investigations using a 3D atmospheric input with a much finer grid spacing to check whether the principal results of papers II and III continue to hold.

\section{Model ingredients}\label{sec:fluxsheet_model}
\subsection{Model atmosphere}\label{sec:mhdfe23_model_atmos}

The input model atmosphere for our calculations was taken from a \mbox{3-D} radiation MHD simulation calculated with the MURaM code \citep{voegleretal2005}. The employed cube is the same as the one used in \citet{smithaetal2020a, smithaetal2020b}. The cube was composed of $1024 \times 1024 \times 256$ voxels spanning a geometrical range of approximately $6\ {\rm Mm} \times 6\ {\rm Mm} \times 1.5\ {\rm Mm}$. The initial seed field was  homogeneous and vertical, with mixed polarity in a 2x2 chequerboard pattern,{ i.e. the left upper and right lower quadrants of the whole domain have been filled with a negative, the other two with a positive homogeneous vertical field of $200$\,G each}. This simulation, with periodic horizontal boundaries and vertical field at the top boundary, with in and outflows allowed, but no passage of magnetic flux  at the bottom boundary, was allowed to run for nearly 30 minutes of solar time after the introduction of the initial field before a snapshot was chosen for analysis. {The magnetic flux is thus conserved during the simulations.}

The 2x2 chequerboard pattern is still visible in the final snapshot from the map of its vertical magnetic field at $\log (\tau_c^{500})=0.0$ shown in Figure~\ref{Figure:fig1}. {When the map is divided into quadrants, the top-left and the bottom-right parts carry mostly negative polarity fields while in the remaining two parts those with positive polarity prevail, and hence the 2x2 chequerboard pattern.} For the RT calculations, the optically thick part at the bottom was cut as in paper III, ending up with a $1024 \times 1024 \times 150$ data cube spanning a height range from approximately 430 km below the average $\tau_c^{500}=1$ (hereafter referred to as $\tau_c$) level to 740 km above it. The smallest $\tau_c$ value at the bottom of the cube was $\approx 38$, while the average value at that depth amounts to almost $1000$.

\begin{figure}
\includegraphics[width=0.45\textwidth]{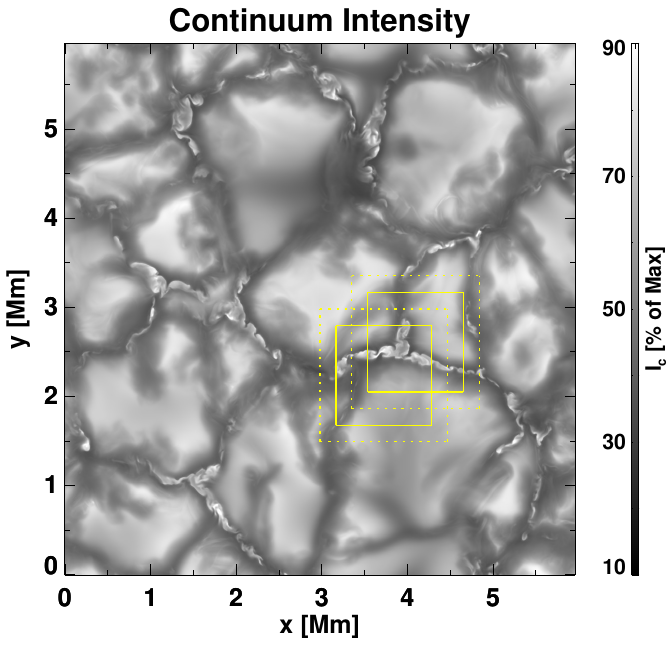}
 \caption{Continuum intensity at a wavelength close to the $525$\  nm lines. The squares indicate the two selected subdomains which were calculated in full \mbox{3-D} NLTE. The bigger dashed squares indicate the domain used for calculations while the smaller solid squares within the dashed squares represent the areas which were used for our investigation.} 
 \label{Figure:fig2}
\end{figure}

Figure~\ref{Figure:fig2} displays the continuum intensity, $I_c$, (close in wavelength to 525~nm) from the snapshot chosen for analysis. As the available computing resources were not large enough to make a true \mbox{3-D} NLTE calculation with the full atmospheric model, we had to choose between lowering the resolution for the 3D RT (as has typically been done in the literature), or restricting the spectral synthesis to only a part of the data cube. As the high resolution of the simulation is central to the aims of this paper, we have selected a representative sub-domain ---containing FT- and FS-like structures --- for which we performed a full \mbox{3-D} NLTE calculation.  Figure~\ref{Figure:fig2}  indicates the location of these two sub-sections of the full box in which 3D RT was carried out (yellow boxes).

\begin{figure*}[htbp]
\begin{center}
 \includegraphics[width=0.82\textwidth]{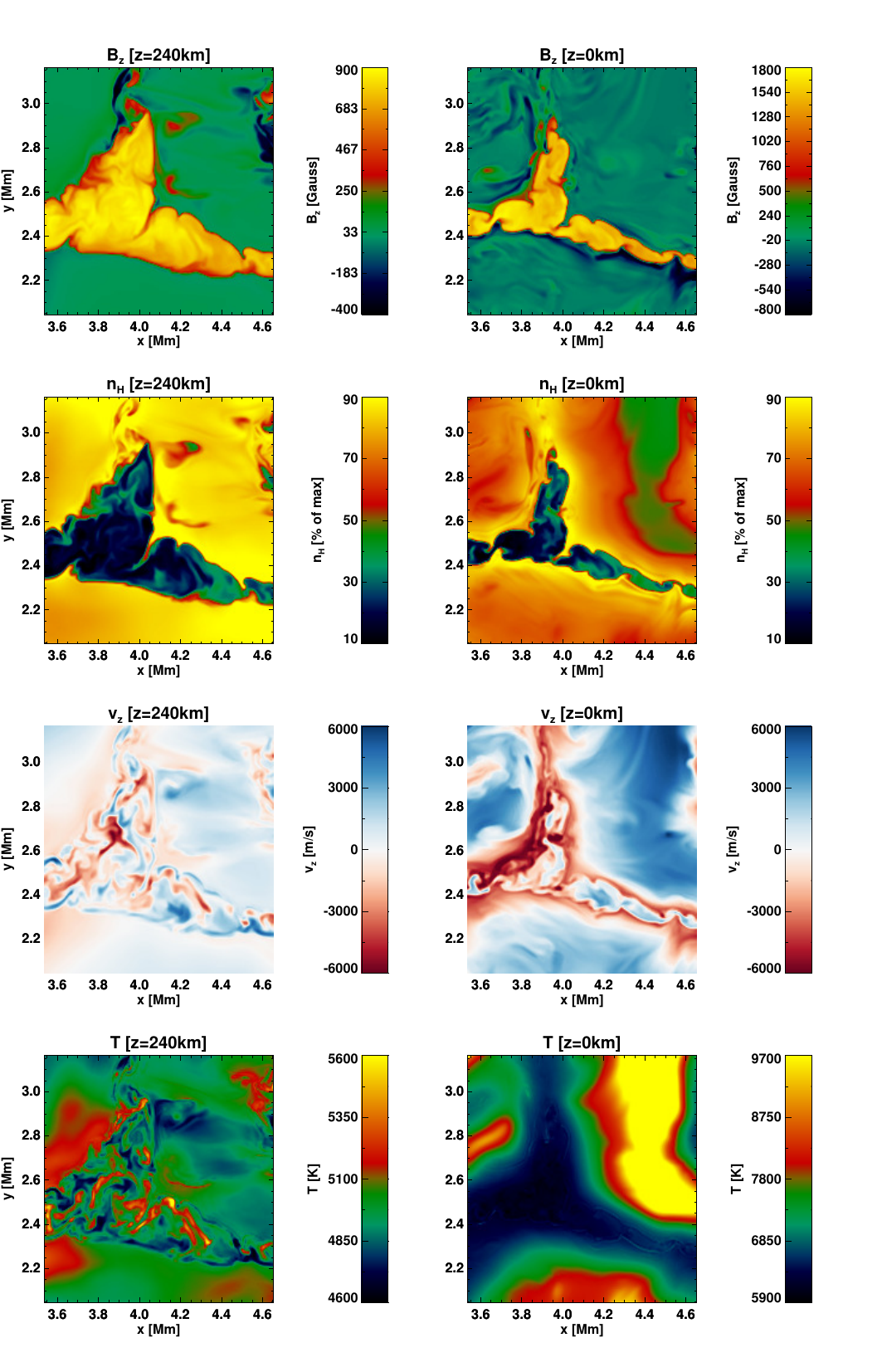}
 \caption{Main atmospheric input parameters of the model sub-domain used in this work at heights $z=0$\,km (i.e. at the height of average $\tau_c=1$; panels on the right) and $z=240$\,km (the approximate line core formation region of the investigated lines; panels on the left). Plotted are, from top to bottom, vertical component of the magnetic field, $B_z$, hydrogen number density, $n_{\rm H}$, vertical component of the velocity, $v_z$, and the temperature, $T$.}
 \label{Figure:fig3}
 \end{center}
\end{figure*}

Some important atmospheric parameters for the selected subdomain at the same geometric heights as in papers II and III, i.e. at $z=0$ km (average height of $\tau_c=1$) and $z=240$ km (approximate average formation height of the cores of the considered lines) are given in Figure~\ref{Figure:fig3}. The average height of $\tau_c=1$ in the selected subdomain is equal to that for the whole atmosphere. The selected subdomain is located at the boundary of three granules. At $z=0$, each of the three intergranular lanes contains strong magnetic flux sheets of up to $1.8$ kG. The density increases towards the boundary of the granules forming thick and relatively cool downflowing walls close to the flux elements. In the center of the magnetic elements the downflows are often reduced or even inverted. The lower right intergranular lane even shows strong upflows in the center of the FS. At $z=240$ km the situation is much more complex when the temperature and velocities are considered. Especially in the intergranular lanes, which are much broader at that height, the atmospheric parameters vary strongly on a very short scale of a few $10$ km.

\subsection{RT-Calculations}\label{sec:mhdfe23_model_RT}
As in the previous papers of this series, we used the code RH \citep{uitenbroek2000, uitenbroek2001} with adaptions as mentioned in papers I and II. The same atomic model (23 levels, 33 lines, and $\approx 1300$ wavelength points) was used as in paper III. To investigate the influence of horizontal radiative transfer and NLTE effects separately, we again calculated the line spectra each in LTE, \mbox{1-D} NLTE, and in true \mbox{3-D} NLTE. For details we refer the reader to these papers.

Cutting a domain from the full atmospheric box violates the assumption of periodic boundary conditions in the \mbox{3-D} version of the RH code, i.e., at the boundaries of the sub-models the atmospheric quantities are not continuous anymore. At a point close to the boundary the local state of the atoms can then erroneously be influenced by a region close to the opposite boundary, i.e. from a part of the atmosphere reasonably far away, definitely further than the horizontal photon mean free path. To investigate the influence of the discontinuity, we selected two subdomains, each spanning $256 \times 256$ grid points but one shifted by 64 grid points along both the $x$ and $y$ coordinate against the other. For both subdomains a full \mbox{3-D} NLTE calculation was performed. These two selected sub-domains are marked by the dashed squares in Figure~\ref{Figure:fig2}.

We choose the Equivalent width (EW) of the $630.15$ nm line to check the influence of the non-periodic boundary conditions on the results of the \mbox{3-D} RT at the boundaries. The EW calculated in any two \mbox{3-D} runs should be equal if the effects at the non-periodic boundaries are negligible. By visual inspection of the profiles at the boundaries of each sub-domain, we made sure that they have a similar shape in both runs, ruling out the possibility that the EW values are not equal by chance due to different shapes of the profiles. 

\begin{figure}
\includegraphics{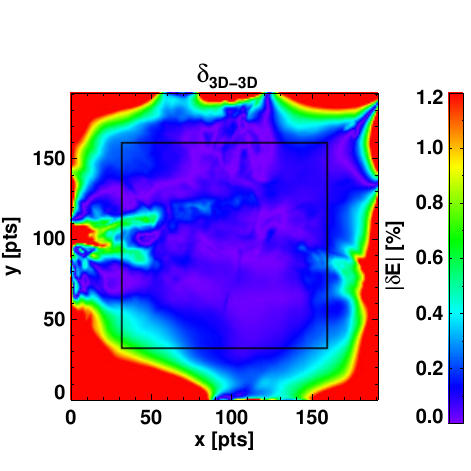}
 \caption{Spatial distribution of the relative differences between the EW values of the Fe I 630.15 nm spectral line calculated in the two spatially shifted sub-domains in \mbox{3-D} NLTE. The colors denote the deviation in the EW between the 2 runs. The rectangle indicates the area chosen for our investigation, which lies completely inside the region with less than $1\%$ difference. } 
 \label{Figure:fig4}
\end{figure}

Figure~\ref{Figure:fig4} presents the relative differences of the EW values for the 630.15 nm line in pixels common to both sub-domains, each calculated with one of the \mbox{3-D} runs ($\delta E_{3D-3D}$) where
\begin{equation}\label{eq:abbrev_ew}
\delta E_{M_i-M_j}
\,\,=\,\,
\frac{EW^{M_i}-EW^{M_j}}{EW^{M_j}},
\end{equation}
with each $M_i$, and $M_j$ referring to one out of the three RT calculation methods (3D, 1D, LTE), or, in the above case, the two spatially shifted \mbox{3-D} runs.

As expected, the effects are largest close to the boundary and decrease inwards. At the very boundary, at some selected locations, the difference between the two EWs is as high as $30\%$. It decreases very rapidly towards the inner part of the common spatial area. By cutting 32 points, corresponding to slightly less than $200$ km at each boundary, we have an agreement in EW which is clearly better than one percent in the whole area. In the following, all results are derived in the area left after the removal of these outer 32 spatial points.

\section{Results}\label{sec:mhdfe23_results}

In paper III the horizontal size of a voxel in the MHD input atmosphere was roughly $21$ km. The cube used in this investigation has a horizontal voxel size of slightly below $6$ km surpassing all \mbox{3-D} NLTE RT calculations performed so far (to our knowledge). This enables us to investigate effects of horizontal RT at a spatial scale clearly below the resolution limit of the new generation telescopes. In the following, we will investigate the strength of horizontal RT effects on spatial scales smaller than the horizontal photon path of roughly $100$ km in the photosphere.

\subsection{Horizontal radiation transport in a highly resolved flux element}\label{sec:mhdfe23_results_continuum}
\begin{figure}
\includegraphics[width=0.5\textwidth]{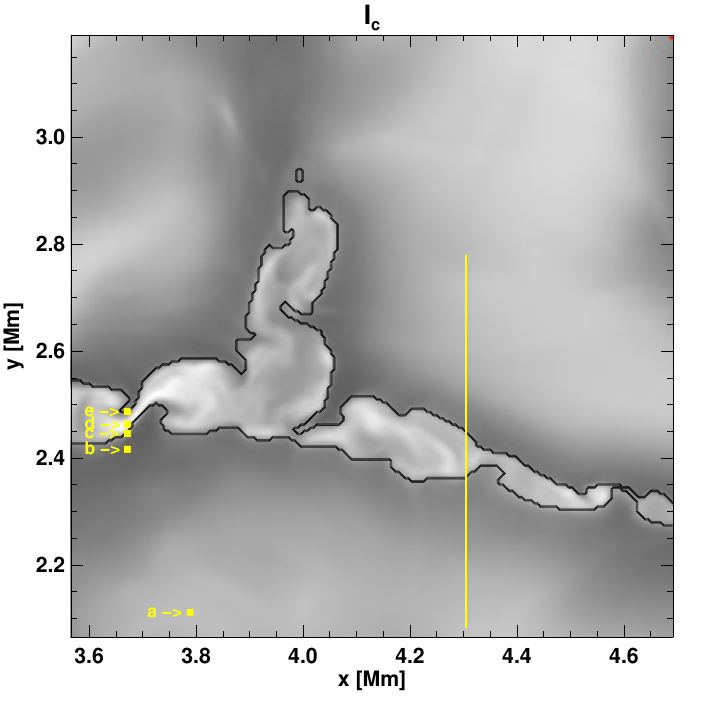}
 \caption{Enlarged section ($\approx 1$ Mm$ \times 1$ Mm) of the continuum intensity image (Figure~\ref{Figure:fig2}) showing the part of the atmosphere where the \mbox{3-D} NLTE calculation was performed. The yellow line and the five yellow dots (a to e) indicate selected positions used in the text for further analysis. The yellow dots were drawn larger ($2 \times 2$) than in the actual image for better visibility. The black contours indicate the boundary of the magnetic element corresponding roughly to $B_z(z=0) = 1200$\,G.}
 \label{Figure:fig5}
\end{figure}

\begin{figure}
\includegraphics[width=0.45\textwidth]{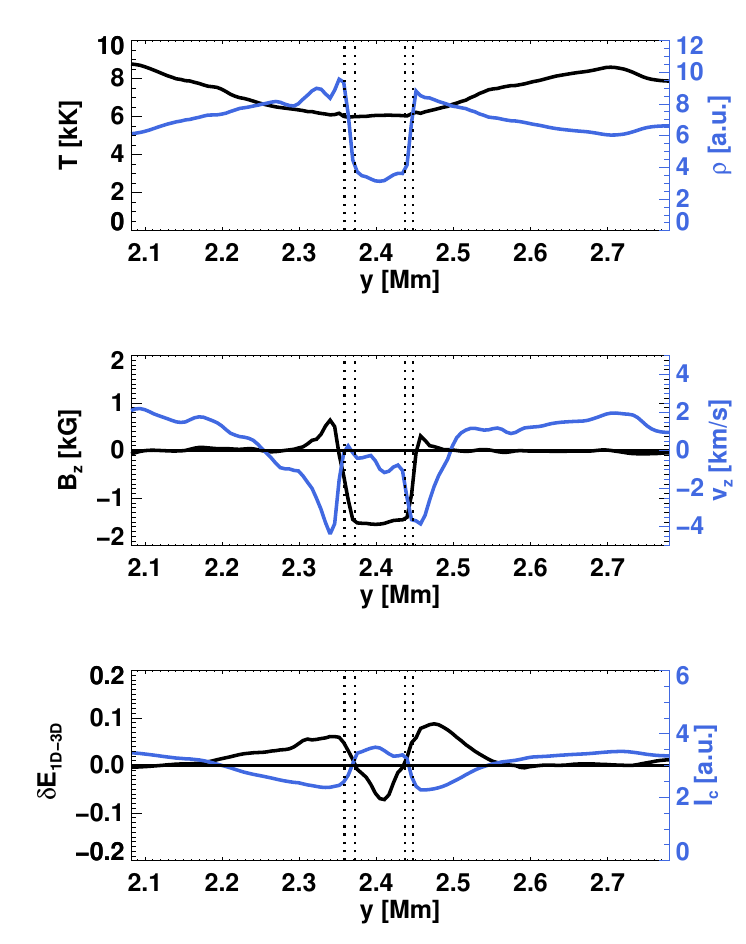}
 \caption{Selected quantities along a cut through a flux sheet indicated in Figure~\ref{Figure:fig5}. Top panel:  Temperature (black) and density (blue); middle panel: vertical magnetic field (black) and vertical velocity (blue, negative values denote downflows); bottom panel: $\delta E^{1D-3D}$ (black) and continuum intensity $I_c$ (blue). All the quantities are taken at geometrical height $z=0$ km. The vertical dotted lines indicate the boundary region of the flux element over which the density decreases and the magnetic field increases.} 
 \label{Figure:fig6}
\end{figure}

As in papers I and III of this series, we first selected a cut through a flux element along which the influence of 1-D NLTE and horizontal RT (3-D NLTE) is studied. Figure~\ref{Figure:fig5} presents the spatial distribution of the continuum intensity $I_c$ in the part of the atmosphere where the \mbox{3-D} NLTE calculation was performed (see also the upper right solid square in Figure~\ref{Figure:fig2}, and Figure~\ref{Figure:fig3}). The yellow line indicates the position of the cut, which extends from one granular element to another, cutting a relatively small ($\approx 80$ km diameter) but typical FS containing a strong vertical magnetic field  of $B_z(\tau_c=1)>1.5$ kG.

Figure~\ref{Figure:fig6} shows different quantities along the cut at the height $z=0$ km (the average $\tau_c=1$ level). In the top panel, the temperature and density profiles along the cut are shown. While the density increases slowly as we approach the FS and abruptly falls within the magnetic part of the FS, the lowering of the temperature is more continuous and much broader than the flux element itself. On the basis of the temperature curve alone, the spatial location of the FS could not even be identified. This is in good agreement with the scenario found in paper III for MURaM simulations with lower resolution, but does not conform to those of paper I and \citet{stenholmstenflo1977,stenholmstenflo1978,brulsvdluehe2001}, where idealized models with a strong temperature gradient at the FS/FT boundary were used. {This demonstrates that magnetic diagnostics are much better suited to identifying structures than the intensity.}

The middle panel presents the variation of the vertical magnetic field $B_z$ and the vertical velocity $v_z$ (positive values denote upflows). In the FS, $B_z$ reaches approximately $1.5$ kG, whereas in the granules it typically lies below 100 G (although it reaches nearly 300 G at one location). In the small region with intense downflows just outside the FS, one observes a field enhancement of up to $500$ G but with opposite polarity. {The opposite polarity flux is associated with the kG flux tube. It can be brought to the surface through flux emergence due to turbulent motions \citep{chitta2019}. Another possibility is that the opposite polarity flux is brought down from the canopy by the downflows \citep{pietarila2011}.  Both of these are due to small-scale intergranular lane turbulence distorting the field lines at the edge of the tubes.} This corresponds to the opposite polarity fields deduced by \citet{buehleretal2015} from Hinode data. As expected, the vertical velocity points upwards in the granules and strongly downwards close to the flux element. In the inner parts of the flux element, $v_z$ is much smaller. The strong downflows are located in areas of weak or vanishing $B_z$. In the inner parts of the selected FS, the velocity is not homogeneous. At some $x$ locations of the FS, weak to medium-strength downflows occur while at others, upflows may be observed (see also Figure~\ref{Figure:fig3}).

In the lowest panel of Figure~\ref{Figure:fig6}, the equivalent width ratio $\delta E_{1D-3D}$ is plotted. It serves as a measure of line weakening ($\delta E>0$) or strengthening ($\delta E<0$) by horizontal RT. The line weakening observed in the \mbox{3-D} NLTE calculation as described for the first time by \cite{stenholmstenflo1977} is only present in the boundary area \emph{outside} the magnetic element. Thanks to the higher resolution of the simulations used here, it is even more obvious than in paper III, that there is no line weakening in the inner parts of the FS. Contrary to the findings from idealized flux tube models \citep[e.g.][paper I]{stenholmstenflo1977, brulsvdluehe2001} a distinct line strengthening is found in the center of the flux element due to the cooler surroundings \citep[see paper III and][]{smithaetal2020a}. Although along the cut, the temperature increases away from the center of the FS (top panel in Figure~\ref{Figure:fig6}), the surroundings of the FS are on average cooler compared to the center as seen in Figure~\ref{Figure:fig3}. Due to horizontal transfer these cooler surroundings result in an overall weakening of the line profile.
The blue curve in the lowest panel of Figure~\ref{Figure:fig6} represents the continuum intensity $I_c$ at $\approx 525$ nm.  The variation in $I_c$ is opposite to that of $\delta E_{1D-3D}$ confirming the findings of papers II and III, where we found that in areas with higher $I_c$, the horizontal effects tend to strengthen the spectral lines whereas in areas with lower $I_c$, horizontal transfer weakens the spectral lines.

\subsection{Are the walls resolved?}
\label{sec:mhdfe23_results_wall_thickness}
The thickness and the temperature profile of the wall of a magnetic element are central factors determining the line weakening/strenghtening of \ion{Fe}{i} lines calculated in \mbox{3-D} NLTE (see also paper I where we investigated the influence of wall thickness). An open question in paper III was whether a better spatial resolution would give rise to a different picture of this situation. There, the boundary of a flux element -- defined by the drop in density from $90\%$ to $10$\% extended over two to three voxels corresponding to a spatial scale of approximately $50$ km. With our current high-resolution atmosphere the thickness of the wall still extends approximately over three voxels, corresponding now to less than $20$ km. Whether the wall is fully resolved or not still remains unanswered. Presumably the thickness of three voxels is limited by the MHD modelling rather than by any other influencing factors. It is quite possible that with an even higher resolution in the MHD calculation, the wall would be even thinner. One could be tempted to believe that such a thinner wall could then decrease the separation of the cooler inner parts of the FS from the outer hotter surroundings as is the case in FT cartoon models and therefore increasing the UV irradiation into the flux element.
However, if we look at Figure~\ref{Figure:fig6}, we find that the temperature gradient is not determined by the density drop nor by the field strength enhancement. It spreads even far beyond the width of the downflow area outside the wall.
{We conclude that, even when the wall is not resolved, our results for the temperature profile, and, therefore, the observed horizontal RT effects, will not change qualitatively if the resolution is increased further. However, quantitative changes are of course possible. }

The downflow area is much broader than three voxels and therefore likely to be fully resolved. In paper III, the downflows spread over the whole flux element and beyond, whereas they are now clearly concentrated in the immediate surroundings of the magnetic element. Thus we can conclude that the fundamental picture is not likely to change with an even further increase in the spatial resolution of the simulation box. Owing to the better localization of the wall and especially the downflows, one could speculate that the strength of horizontal RT effects should increase with increasing resolution, at least to some extent. However, if we compare the strength of the line weakening/strengthening in the present atmosphere with that of paper III, the effects are even slightly smaller in the present, high-resolution case.

\subsection{Variation of transfer effects on smaller scales}
\label{sec:mhdfe23_results_wall_thickness}
\begin{figure*}
\includegraphics[width=\textwidth]{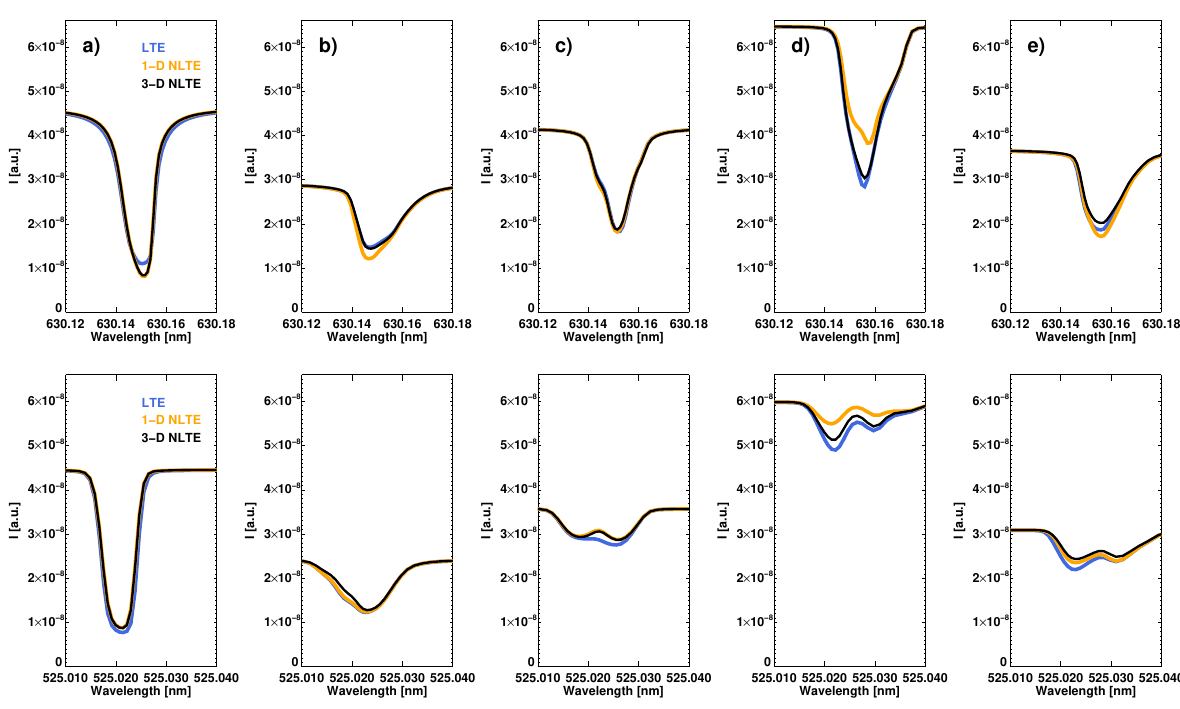}
 \caption{Intensity profiles of the $630.15$ nm (upper row) and the $525.02$ nm (lower row) lines at five selected spatial positions as indicated in Figure~\ref{Figure:fig5}. Ordering: a) location with lowest y coordinate (granule) to  e) location with largest y coordinate (at the upper boundary of the flux sheet). Legends explaining the colours of the curves are given in panels a).} 
 \label{Figure:fig7}
\end{figure*}

We now investigate the spatial scales on which 1-DNLTE and horizontal RT effects may vary. Whether the photon mean free path $\lambda$ could serve as a lower limit for the scales on which the 3-D NLTE effects can be observed has been debated for a long time. Based on 3D-NLTE calculations of the the chromospheric \ion{Ca}{I} 422.7 nm line, \citet{judgeetal2015} argued that the amount of scattering in an observed line is the main factor controlling the extent of smearing due to horizontal RT effects. Since the photospheric lines are in general less scattering than the chromospheric lines, it should {in principle} be possible to find variations at the photosphere on spatial scales below $\lambda$ because the temperature, the density, and the magnetic field, respectively, vary on scales much smaller than $\lambda$ \citep[e.g.][see also the width of the FS walls in Fig.~6]{schuessler1986}.

We will show in this section, that 3-D NLTE effects in the iron lines considered here may vary on scales much smaller than $\lambda$. We selected five locations in our atmosphere represented by the yellow dots in Figure~\ref{Figure:fig5}. From the bottom to the top: a) is located in the middle of a granule, far away from any magnetic elements, b) lies in a downflow region with low $I_c$ adjacent to a FS, c) is situated at the boundary of the FS, between bright and dark $I_c$ regions, d) lies in the middle of the FS (high $I_c$) and e), again, is located at the boundary but on the other side of the FS. We note that locations b) to e) lie close together, each less than $25$ km from its neighbors.

Figure~\ref{Figure:fig7} shows (typical) line profiles of the $630.15$ nm and the $525.02$ nm lines at the selected positions. In the granule (a), the profiles for \mbox{1-D} NLTE and \mbox{3-D} NLTE virtually coincide. No major effects due to horizontal RT occur, as expected from the relatively homogeneous atmospheric conditions.  The LTE profiles differ slightly from their NLTE counterparts, the LTE line core intensity in the $630.15$ nm line being slightly higher, the one in the $525.02$ nm line slightly lower. At position b) in the downflow area close to the FS, the \mbox{3-D} NLTE lines are weakened relative to their \mbox{1-D} NLTE counterparts owing to strong UV irradiation from the hot area in the granule {at the bottom}. We note that this location corresponds best to the situation given in the cartoon models of \citet{stenholmstenflo1977, stenholmstenflo1978, brulsvdluehe2001} and  paper I. At position c), however, the differences between the \mbox{3-D} {and} \mbox{1-D} NLTE line intensities disappear. The place marks the reversal point of the effects between locations b) and d), respectively. In the middle of the FS, at location d) a considerable strengthening of the \mbox{3-d} NLTE lines is observed (as also found in paper III). {Position e) at the other boundary is much more complex as it lies close to the FS boundary where the FS folds on itself. Hence the profiles at e) are influenced by a complex mix of both hotter and cooler surroundings. {The line core intensities in both the $630.15$ nm line and the $525.02$ nm line are weakened by the horizontal transfer effects}.
Note that the $630.15$ nm line is affected by both scattering as well as UV overionization effects while for the $525.02$ nm line, UV overionization is the dominant NLTE mechanism \citep[see paper II, III,][]{smithaetal2020a, smithaetal2020b, smithaetal2023}. Hence at points a) to e),  the 1-D NLTE profile of the $630.15$\,nm line can either be stronger or weaker than the LTE profile whereas for the $525.02$\, nm line, the 1-D NLTE profile is always weaker than the LTE profile.

\begin{figure}
\includegraphics[width=0.45\textwidth]{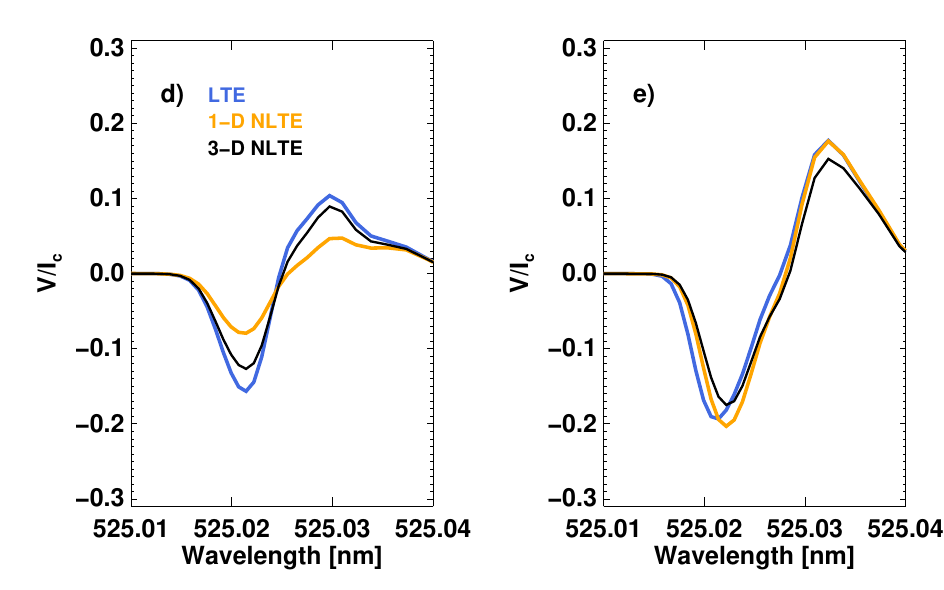}
 \caption{Sample profiles of Stokes $V/I_c$ of the $525.02$ nm line at the two locations d) and e).{ See Figures~\ref{Figure:fig5} and \ref{Figure:fig7} for an explanation of the locations}.}
 \label{Figure:fig8}
\end{figure}

These sample profiles confirm the findings of paper III. Furthermore, they show that variations due to horizontal RT may be observed on rather small spatial scales. The \mbox{3-D} NLTE line strengthening at position d) and the line weakening at position e) lie only 4 voxels apart from each other, i.e., the spatial distance between d) and e) is less than $25$ km. Similarly, locations b) and d), lie $45$ km apart. These are typical distances on which line strengthening reverts to line weakening as can be seen also from the $\delta E^{1D-3D}$ curve in the bottom panel of Figure~\ref{Figure:fig6} which displays the effect of horizontal RT along the cut indicated in Fig.~5. There, $\delta E^{1D-3D}$ also varies strongly within approximately $50$ km. Over a given region, the larger the local horizontal gradients in temperature and intensity, the stronger are the variations in profiles due to horizontal RT \citep{carlssonetal2004}.

Figure~\ref{Figure:fig8} shows two sample profiles of Stokes $V/I_c$ at locations d) and e). The same effect as in the  $I$ profiles (see Figure~\ref{Figure:fig7}) can be found. The strength of the NLTE/RT effects are of similar magnitude as in Stokes $I$, a finding which has been stated already in \citet{stenholmstenflo1978} based on a simple  FT cartoon model. In some cases, we observe that the $V/I_c$ maxima are slightly shifted in wavelength relative to the Stokes $I$ profile.
This is most probably owing to the height dependence of the vertical velocity, as the Stokes $V/I_c$ profiles are  formed deeper in the atmosphere than the Stokes $I$ line core. Therefore, the Stokes $V/I_c$ profiles probe a slightly different height regime.

\subsection{Contrasts and spatial smearing}
\label{sec:mhdfe23_results_contrasts}

\begin{figure*}
\centering
\includegraphics[width=\textwidth]{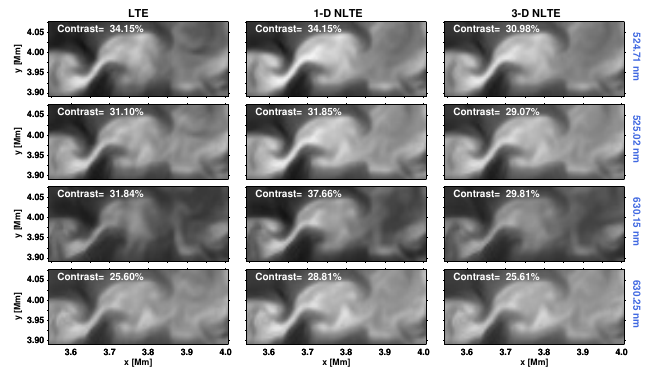}
 \caption{Comparison of RMS contrasts over a small region of the magnetic element for the three computation methods: LTE (first column), 1-D NLTE (second column) and 3-D NLTE (third column). In each row, we show the intensity images at the rest wavelength of the respective line, indicated in blue. The color scale is the same as in Figure~\ref{Figure:fig10}.} 
 \label{Figure:fig9}
\end{figure*}

The horizontal radiative transfer can affect the quality of the observed images in two ways, one by reducing the image contrasts \citep{holzreutersolanki2013, holzreutersolanki2015}
and the other by smearing out the spatial structures \citep{judge2020}. In the sections below we discuss these two effects one after the other.

\subsubsection{Effect on contrasts}
In order to understand effects of horizontal transfer on the brightness contrast in the line cores, we investigate how strongly the RMS contrast changes depending on which of the three different computation methods is used (LTE, 1-D NLTE, and 3-D NLTE). The RMS contrast is defined as the standard deviation divided by the mean, expressed in percentage \citep{danilovicetal2008}. In all cases, the RMS contrasts were calculated at the rest wavelengths of the lines. 

A large part of the spatial region chosen for 3-D NLTE computations (Figure~\ref{Figure:fig5}) is covered by granules which appear dark in the line core and have low RMS contrast. As the contrast depends on the area selected, i.e. the structures contained and their sizes, we investigate the influence of the computation methods on different geometrical scales. To better quantify the effect of 3-D NLTE on contrasts, we chose a smaller region approximately corresponding in size to the mean free photon path $\lambda$. The region contains a part of a magnetic element harboring strong density variations.  This region is shown in Figure~\ref{Figure:fig9} for all lines and computation methods with the RMS contrast in each case indicated in the respective panel of Figure~\ref{Figure:fig9}. 

\begin{figure*}
\centering
\includegraphics[width=0.7\textwidth]{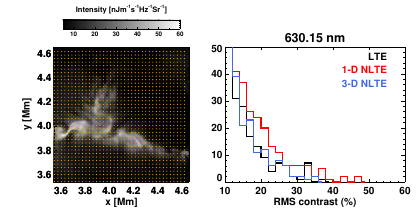}
 \caption{Histogram of RMS contrast of small areas: we divide the full area of the intensity image into a grid of squares each having a size of 5 pixels $\times$ 5 pixels. The grid is indicated as orange dots on the intensity image shown in the left panel. A histogram of the RMS contrasts computed at every square in the grid is shown in the right panel, for the three computation methods: LTE (black), 1-D NLTE (red), and 3-D NLTE (blue). The x-axis is clipped to show only bins with contrasts $\ge 10\%$.} 
 \label{Figure:fig10}
\end{figure*}

Between the 525 nm line pair and the 630 nm line pair, the contrast is largest for the 524.71 nm line because it has a lower excitation potential and is formed deeper in the atmosphere compared to the lines in the 630 nm pair. For the 525 nm pair, the change in contrast going from LTE to 1-D NLTE is quite small while the 630 nm line pair, interestingly, shows an increase in contrast. {Figures 5 and 6 of Paper II demonstrate that the contrast for the lines in $630$ nm pair can either decrease or increase going from LTE to 1-D NLTE depending on whether the contrast is computed at the minimum of the line core intensity or at the rest wavelength of the line. On the other hand, for the $525$ nm pair, the contrast decreases in both cases. Since the contrasts in Figure~\ref{Figure:fig9} are computed at the rest wavelength of the lines, we see an increase in contrast going from LTE to 1-D NLTE for the $630$ nm pair.}
On the other hand, the 3-D NLTE computations in all cases show reduced contrasts relative to 1-D NLTE, with the 630.1 nm line showing the largest reduction. This decrease in contrast in 3-D NLTE computations can be attributed to the horizontal transfer effects. 

A similar analysis was carried out in Paper II, i.e. for the purely hydrodynamic case. There the difference in contrast was larger between LTE and 1-D NLTE than between 3-D NLTE and 1-D NLTE. This could not only be due to differences in the simulations but also that the actual values of the contrasts are dependent on the area considered and the type of features it contains. The is supported by the histogram plotted in Figure~\ref{Figure:fig10} where we divided the entire region into small squares of size 5 pixels $\times$ 5 pixels. In each square, the contrast was determined for the intensities computed in LTE, 1-D NLTE and 3-D NLTE. Then a histogram was created for regions which show contrast $\ge 10\%$. Only the 630.15 nm line was chosen for this analysis since the contrast in this line showed the strongest response to 3-D effects compared to other lines, as seen from Figure~\ref{Figure:fig9}. This is because of all the four lines considered, the 630.15 nm is most affected by scattering \citep[see paper III,][]{smithaetal2020a}, and the stronger the scattering effects in a given line, the greater is the reduction in contrast due to horizontal transfer effects according to \citet{judgeetal2015}.

From Figure~\ref{Figure:fig10}, we can learn how the computation method influences the contrast on very small scales (below 25km). The general trend of a contrast reduction by horizontal RT, i.e. by 3-D NLTE computations, is clearly seen from the histogram. In some regions, the contrast from 1-D NLTE reaches values as high as $50\%$ but for 3-D NLTE the contrast values do not exceed $37\%$. The number of areas showing large contrasts is also strongly reduced for 3-D NLTE. E.g. the number of areas showing contrast values of $\ge 25\%$ is more than halved from approximately $40$ (1-D NLTE) to less than 20 (3-D NLTE). Overall, on very small scales (here, below $25$ km), horizontal transfer effects may degrade the contrast but on a more global scale the influence is much smaller, of course depending on the type of feature considered. As magnetic structures show very strong intensity variations on even smaller scales, it is still possible to resolve the structures --- given the necessary telescope resolution --- though with reduced contrast. This can also be seen from Figure~\ref{Figure:fig9}.

\subsubsection{Spatial smearing}
In addition to reduction in image contrast, horizontal transfer effects  spatially smear structures in the solar atmosphere. {Signatures of spatial smearing due to 3D effects are evident by comparing, for example, the 1-D NLTE and 3-D NLTE maps for the 630.1 nm line in Figure~\ref{Figure:fig9} where we present a blow-up of a small region around the flux tube.}
In observations,  spatial smearing is also introduced by the diffraction limit of the telescope. To compare the two effects, we spatially degraded our computations. Once again, we choose the 630.1nm line  since it is more affected by the 3-D NLTE effects compared to the other lines considered in this study. 

The intensity images from both 1-D NLTE and 3-D NLTE computations were spatially degraded to match the specifications of the Visible Spectro-polarimeter  \citep[ViSP;][]{dewijnetal2022} at DKIST. The degradation was done in two steps: first the intensity was convolved with a Gaussian, mimicking a point spread function (PSF) having a FWHM of 0.07 arcsec corresponding to the theoretical spatial resolution of ViSP at 630 nm\footnote{\protect\url{https://nso.edu/telescopes/dkist/instruments/visp/}}. The convolved profiles were then spatially rebinned to the detector pixel size. No spectral degradation was applied.

In Figure~\ref{Figure:fig11}, we present the intensity images for the 630.1 nm line at its rest wavelength, from both 1-D NLTE and 3-D NLTE computations before and after the two step spatial degradation. The contrast decreases from 1-D NLTE to 3-D NLTE computations irrespective of whether instrumental degradation is applied or not, see Figure~\ref{Figure:fig11}. Surprisingly, the decrease in the global RMS contrast due to the PSF of the telescope ($5\% - 6\%$) is comparable to that caused from horizontal transfer in 3-D NLTE calculations ($5\% - 6\%$). 
Convolution with the PSF of DKIST causes a much stronger spatial smearing compared to that caused by 3-D NLTE effects. The finer structures within the magnetic element are still visible after carrying out 3-D NLTE computations, but are no longer visible after applying the DKIST PSF. Rebinning to detector pixel size, which is nearly five times larger than the grid spacing in the MHD simulations, further degrades the image quality, by reducing the smoothness of the intensity image, but has little effect on the contrast and seemingly also not on the resolution.

\begin{figure}
    \centering
    \includegraphics[width=0.49\textwidth]{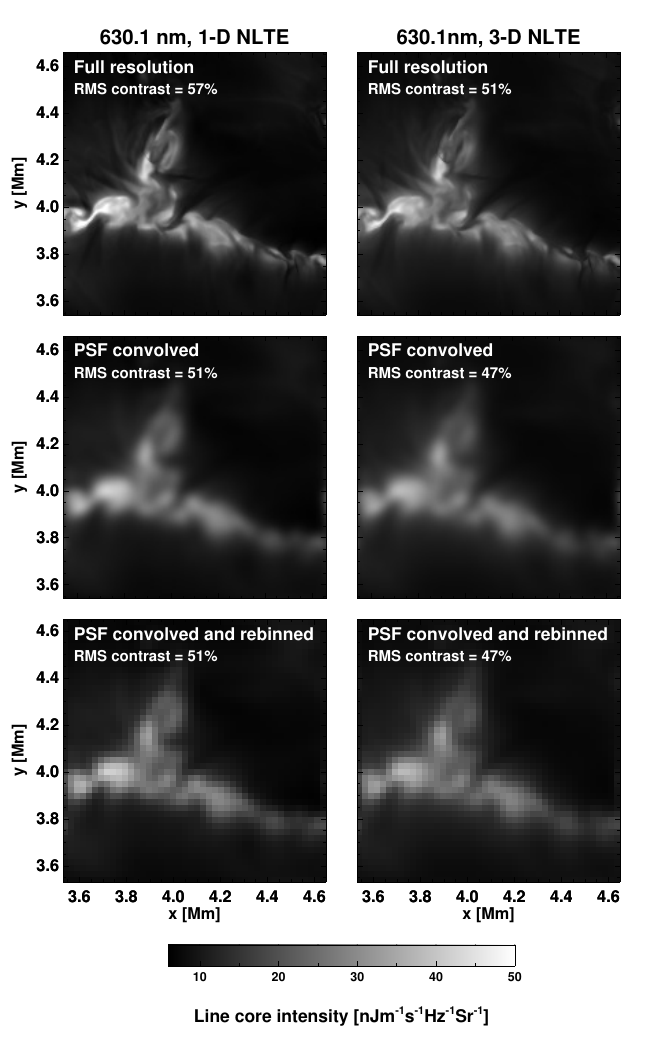}
    \caption{Comparison of intensity maps at the rest wavelength of $630.1$ nm line from 1-D NLTE and 3-D NLTE computations. The top row shows images at full resolution of the MHD cube while for the images in the lower rows a spatial degradation has been applied, first by convolving with a PSF (middle row) and then by rebinning to the detector pixel resolution (bottom row).}
    \label{Figure:fig11}
\end{figure}

\section{Discussion and conclusions}
\label{sec:mhdfe23_discussion}
In the present paper we have investigated how strongly the horizontal transfer of radiation influences the visibility of structures at very small spatial scales in widely used photospheric spectral lines. To this end, we carry out 3D radiative transfer in a high resolution 3D radiation MHD simulation box of a plage region with a horizontal grid spacing of 6\,km, which is three times finer than in our previous studies. This allows us to test also the effects of horizontal transfer on the finest spatial scales that are accessible with the currently largest solar telescope, the DKIST on Maui. 

First we test if the effects introduced by horizontal transfer and found in our Paper III are reproduced at higher resolution. They are, and are even clearer at higher resolution, which leads to sharper boundaries of magnetic elements and thus to a cleaner separation of the gas properties inside and outside of the elements. 

After that we consider intensity contrasts and find that in the line cores the contrast generally decreases from 1-D NLTE to 3-D NLTE. On very small scales below $25$ km, the decrease may be considerable but on larger scales it remains small. Of the four lines considered here, the 630.1\,nm line shows the largest reduction in contrast. This reduction is moderate, between $4\% - 6\%$, at both full resolution and after spatial degradation. 

Purely from the horizontal transfer effects in the photospheric layers, we do not see any evidence of the solar "fog" predicted by \cite{judgeetal2015}. The solar "fog" is mainly caused by scattering and in the photospheric lines analysed by us, the effects of scattering are very small. The 525 nm line pair is much less scattering than the 630 nm pair \citep[see also Table 1 of][]{judgeetal2015}. For the 525 nm pair, UV overionization is the dominant NLTE effect and its source function closely follows the Planck function. The 630 nm pair is affected by both UV overionization and scattering. This explains why the 630.1 nm line, which is the stronger in the pair, shows a bigger reduction in contrast compared to all the other lines. However, the scattering coefficient $(1-\epsilon)$, where $\epsilon$ is the collisional coupling parameter, for the 630 nm line pair is much smaller than for chromospheric lines. 
Even on very small scales, where we find a considerable reduction of the contrast, it is still possible to see the large intensity variations produced by magnetic elements. 
Hence the spatial smearing and reduction in contrast due to horizontal transfer effects resulting in a solar "fog" is of concern only for lines formed in the chromosphere. 

Our investigations of the formation of photospheric iron lines in a high resolution 3D MHD simulation conforms with the findings from 2-D idealized flux sheet models of \citet{stenholmstenflo1977, brulsvdluehe2001} and paper I, in the sense that magnetic elements display intensity variations on scales well below the photospheric photon mean free path. We conclude that the spatial resolution achievable in the studied lines is only slightly degraded by horizontal radiative transfer, even at scales well below the horizontal photon mean free path in a plane-parallel atmosphere. This degradation is clearly smaller than that introduced by the PSF of the telescope and the detector for the new generation of large solar telescopes (DKIST and EST).  Therefore, the degradation due to horizontal RT should not hamper the capabilities of these telescopes to capture fine scale structures smaller than the photon mean free path, at least in the photosphere. 

\begin{acknowledgements}
The authors thank R. Cameron for running the high resolution MURaM simulations. {HNS thanks L.~P.~Chitta and D.~Przybylski for helpful discussions. }This project has received funding from the European Research Council (ERC) under the European Union's Horizon 2020 research and innovation programme (grant agreement No. 101097844 -project WINSUN).
\end{acknowledgements}
%


\bibliographystyle{aa}

\end{document}